\documentclass[%
 reprint,
superscriptaddress,
 amsmath,amssymb,
 aps,
]{revtex4-2}

\usepackage{graphicx}
\usepackage{float}
\usepackage{ulem}
\usepackage{xcolor}
\usepackage{dcolumn}
\usepackage{bm}
\usepackage[T1]{fontenc}
\usepackage[utf8]{inputenc}
\usepackage{bbm}

\newcommand\numberthis{\addtocounter{equation}{1}\tag{\theequation}}
\DeclareUnicodeCharacter{202F}{\,}
\begin{document}

\preprint{APS/123-QED}

\title{Signatures of coherent phonon transport in frequency dependent lattice thermal conductivity}

	\author{{\DJ}or{\dj}e Dangi{\'c}}
	\email{dorde.dangic@ehu.es}
	\affiliation{Fisika Aplikatua Saila, Gipuzkoako Ingeniaritza Eskola, University of the Basque Country (UPV/EHU), 
		Europa Plaza 1, 20018 Donostia/San Sebasti{\'a}n, Spain}
	\affiliation{Centro de F{\'i}sica de Materiales (CFM-MPC), CSIC-UPV/EHU, 
		Manuel de Lardizabal Pasealekua 5, 20018 Donostia/San Sebasti{\'a}n, Spain}

\date{\today}

\begin{abstract}
Thermal transport in highly anharmonic, amorphous, or alloyed materials often deviates from conventional phonon-mediated behavior, exhibiting signatures of coherent transport. While first-principles methods have incorporated this contribution to predict materials with ultra-low lattice thermal conductivity, its direct experimental evidence remains elusive. Here, we propose that the frequency-dependent lattice thermal conductivity, $\kappa(\nu)$, provides a direct signature of coherent transport. Specifically, we show that peaks in $\kappa(\nu)$ arise from the frequency nesting of modes with identical wave vectors. Applying this approach to CuCl, we identify clear signatures of coherent transport in its dynamical lattice thermal conductivity. We revisit the interpretation of thermoreflectance experiments and argue that the conventional understanding breaks down in strongly anharmonic crystals, alloys, and amorphous materials. Finally, we discuss experimental pathways to measure $\kappa(\nu)$, offering a new route to verify coherent contributions in thermal transport.
\end{abstract}

\maketitle

\section{Introduction}
Thermoelectric materials offer a promising route for harvesting waste heat, but their adoption is limited by relatively low heat-to-electricity conversion efficiency. This efficiency is quantified by the thermoelectric figure of merit, which is inversely proportional to the lattice thermal conductivity $\kappa$. Accordingly, the best thermoelectric materials typically exhibit low $\kappa$ values~\cite{GeTeThermoelectrics, PbTeThermoelectrics, SnSeThermoelectrics, BiTeThermoelectrics,MgSbThermoelectrics,LatticeSoftening}. Low lattice thermal conductivity is commonly attributed to factors such as large unit cells~\cite{ClathrateThermoelectric}, strong anharmonicity~\cite{PbTeAnharm,SnSeAnharm}, proximity to structural phase transitions~\cite{Aoife1,Aoife2}, heavy elements~\cite{BiTe}, and lattice imperfections~\cite{GeTeDWs}. First-principles calculations have played a crucial role in identifying and explaining these trends~\cite{GeTemine,SnSeIon,BiTeOlle,PbTeOlle}.

However, in some cases—especially in materials with ultralow $\kappa$—first-principles calculations significantly underestimate the lattice thermal conductivity~\cite{CsPbBr3_kappa}. This discrepancy has been resolved by introducing an additional transport channel, often referred to as the coherent contribution~\cite{twochannel,Simoncelli2019,Isaeva2019,GeTemine,Caldarelli,Simoncelli2022}. This mechanism becomes relevant when phonon modes with the same wave vector have small frequency separations—comparable to the sum of their linewidths. As such, it emerges prominently in materials with strong anharmonicity, where phonons acquire short lifetimes, or in complex, disordered, or amorphous systems with broken translational symmetry.

A landmark success of this approach was the modeling of thermal transport in CsPbBr$_3$ in its low-symmetry phase with four formula units per primitive cell~\cite{Simoncelli2019}. Shortly thereafter, a related formulation was used to successfully model thermal transport of amorphous silicon~\cite{Isaeva2019}. Since then, the coherent transport model has been applied to a variety of materials—complex crystals, alloys, glasses—with encouraging agreement with experiments~\cite{CsPbBr3_newkappa,CsBiICl,TlBiSe2,Fiorentino1,Fiorentino2023,Simoncelli_glasses,alumina,Organic,CsCuSe,Castellano}. Yet, a critical limitation remains: because experiments typically measure only the total thermal conductivity, they cannot directly isolate or confirm the presence of the coherent contribution~\cite{Cahill_comment}.

In this work, we address this gap by proposing a direct experimental probe of coherent heat transport. Using the Green-Kubo formalism, we derive expressions for the frequency-dependent lattice thermal conductivity $\kappa(\nu)$, explicitly including the coherent contribution. This frequency-resolved quantity provides access to average phonon lifetimes and—crucially—reveals spectral features that arise solely from coherent processes, specifically from the nesting of phonon modes with the same wave vector. We apply this methodology to CuCl, a simple but highly anharmonic material, and identify clear signatures of coherent transport at frequencies below 1 THz. Finally, we discuss the experimental feasibility of detecting $\kappa(\nu)$ and its implications for validating the coherent transport paradigm.

\section{Results}
In the case of a uniform heat source (that is, when the heat flux ($\mathbf{J}$) has no spatial dependence) a simplified form of Fourier’s law holds~\cite{heatwaves}:
\begin{align*}
    \mathbf{J}(t) &= -\int_0^{\infty} \kappa(t - t')\, \nabla T(t')\, dt', \\
    \mathbf{J}(\nu) &= -\kappa(\nu)\, \nabla T(\nu). \numberthis \label{eq:fourier_law}
\end{align*}
The second line is simply the Fourier transform of the first. In the time-independent limit, one recovers the more commonly cited form of Fourier’s law with the static thermal conductivity
\[
\kappa = \int_0^{\infty} \kappa(t')\, dt' = \kappa(\nu \rightarrow 0).
\]
To investigate the dynamical lattice thermal conductivity \(\kappa(\nu)\), we begin with the expression for its real part, \(\kappa^{x,y}(\nu)\), derived within the Green–Kubo formalism (see Ref.~\cite{GK-main} for details):
{\footnotesize
\begin{align*}
    &\kappa ^{x,y} (\nu) = \frac{\pi\beta ^2k_{B}}{NV}\sum _{\mathbf{q},j,j'}v^{x}_{\mathbf{q},j,j'}v^{y*}_{\mathbf{q},j,j'}\omega _{\mathbf{q},j}\omega _{\mathbf{q},j'}\frac{e^{\beta\nu} - 1}{\beta\nu}\times \numberthis \label{eq:dyn_kappa}\\
    &\times\int_{-\infty}^{\infty}\textrm{d}\Omega (1+\frac{\Omega}{\Omega + \nu})\frac{e^{\beta\Omega}}{\left(e^{\beta\Omega} - 1\right)\left(e^{\beta(\Omega + \nu)} - 1\right)}\sigma _{\mathbf{q},j'}(\Omega)\sigma_{\mathbf{q},j}(\Omega + \nu). 
\end{align*}}%

Here, $\mathbf{v}_{\mathbf{q},j,j'}$ are the generalized phonon group velocities, $\omega _{\mathbf{q},j}$ are the phonon frequencies, and $\sigma _{\mathbf{q},j}(\Omega)$ are the phonon spectral functions for wave vector $\mathbf{q}$ and mode index $j$. The variable $\nu$ is the driving frequency (frequency of the heat source), and $\beta = 1/k_BT$ is the inverse temperature. Equation~\ref{eq:dyn_kappa} accounts for both the diagonal (particle-like) contribution ($j = j'$) and the coherent (wave-like) contribution ($j \neq j'$). The integral spans the full spectral lineshape, making the expression applicable even in strongly anharmonic regimes where phonon lifetimes and energies are not sharply defined. The imaginary part of $\kappa(\nu)$ can be obtained via a Kramers-Kronig transformation of its real part. 

While Eq.~\ref{eq:dyn_kappa} is general, its physical interpretation is not always intuitive—particularly from a phonon-centric perspective. To clarify its structure, we now consider the perturbative limit, limit of weak anharmonicity. In this regime, the phonon spectral function $\sigma_{\mathbf{q},j}(\Omega)$ can be approximated by a Lorentzian centered at $\omega_{\mathbf{q},j}$ with width $\Gamma_{\mathbf{q},j}$, where $\Gamma_{\mathbf{q},j} = 1/(2\tau_{\mathbf{q},j})$ is related to the phonon lifetime $\tau_{\mathbf{q},j}$. This simplification facilitates analytic insights into the coherent versus particle contributions to thermal transport.

We first consider the perturbative limit for the diagonal case $j = j'$ (corresponding to particle-like transport). In this regime, the dynamical lattice thermal conductivity reduces to:

\begin{align*}
    \kappa ^{x,y} (\nu) &= \frac{1}{NV}\sum_{\mathbf{q},j}\kappa^{x,y}_{\mathbf{q},j}\frac{1}{\nu^2\tau_{\mathbf{q},j}^2 + 1}. \numberthis \label{eq:pert_diag}
\end{align*}

where $\kappa^{x,y}_{\mathbf{q},j}$ is the contribution of the $(\mathbf{q},j)$ phonon mode to the static thermal conductivity. This expression matches the results obtained via the linearized Boltzmann transport equation under the relaxation time approximation~\cite{AC_Chaput, AC_lindsey, AC_Volz}.

As expected, Eq.~\ref{eq:pert_diag} reproduces the static limit of the Boltzmann equation as $\nu \rightarrow 0$. With increasing $\nu$, $\kappa(\nu)$ decreases monotonically—a behavior observed in time-domain thermoreflectance (TDTR) experiments. Interestingly, this decay is faster for phonons with longer lifetimes. In fact, the contribution of a given mode falls to half its static value at $\nu = 1/\tau_{\mathbf{q},j}$. This implies that time-resolved measurements are more accurate in low-thermal-conductivity materials, where $\kappa(\nu)$ remains closer to its static value. Experimentally, the suppression of $\kappa(\nu)$ with frequency is often attributed to a reduced thermal penetration depth from a modulated heat source~\cite{Cahill_kappa_nu, Regner2013}, which limits the contributions of long mean free path phonons. In contrast, Eq.~\ref{eq:dyn_kappa} provides a microscopic interpretation: the decay arises from phonon decoherence induced by the finite frequency of the heat source.

Next, we turn to the off-diagonal (coherent) contribution, corresponding to $j \ne j'$. In the perturbative limit, this contribution takes the form (see Supplementary Material for derivation and a discussion of approximations):

\begin{widetext}
\begin{align*}
    \kappa ^{x,y} (\nu) =  &\frac{\beta ^2k_{B}}{4NV}\sum _{\mathbf{q},j>j'}v^{x}_{\mathbf{q},j,j'}v^{y*}_{\mathbf{q},j,j'}\frac{e^{\beta(\omega _{\mathbf{q},j} - \omega _{\mathbf{q},j'})} - 1}{\beta(\omega _{\mathbf{q},j} - \omega _{\mathbf{q},j'})}\frac{e^{\beta\omega _{\mathbf{q},j'}}(\omega _{\mathbf{q},j} + \omega _{\mathbf{q},j'})^2}{(e^{\beta\omega _{\mathbf{q},j}} - 1)(e^{\beta\omega _{\mathbf{q},j'}} - 1)}\times \\ &\left(\frac{\Gamma _{\mathbf{q},j} + \Gamma _{\mathbf{q},j'}}{(\nu + \omega _{\mathbf{q},j} - \omega_{\mathbf{q},j'})^2+(\Gamma _{\mathbf{q},j} + \Gamma _{\mathbf{q},j'})^2} + \frac{
    \Gamma _{\mathbf{q},j} + \Gamma _{\mathbf{q},j'}}{(-\nu + \omega _{\mathbf{q},j} - \omega_{\mathbf{q},j'})^2+(\Gamma _{\mathbf{q},j} + \Gamma _{\mathbf{q},j'})^2}\right). \numberthis \label{coherent_dyn_kappa}
\end{align*}
\end{widetext}

In addition to this resonant term, a smaller antiresonant component emerges at higher frequencies (see Supplementary Material), but the resonant contribution dominates in the low- and mid-frequency regime. Crucially, the resonant term peaks when the driving frequency $\nu$ matches the frequency difference between two phonon modes with the same wave vector $\mathbf{q}$, or when this difference is comparable to the sum of their linewidths. This implies that, unlike the monotonic decay of the diagonal contribution, the coherent part of $\kappa(\nu)$ can exhibit distinct non-monotonic spectral features that directly reflect mode coherence—analogous to resonant structures in the optical conductivity of electrons.

This observation is key: the appearance of peaks in $\kappa(\nu)$ provides direct evidence of the coherent transport mechanism. To date, such contributions have only been inferred indirectly, via discrepancies between experiment and traditional Boltzmann-based predictions. Our formalism shows that measuring the frequency-dependent structure of $\kappa(\nu)$ offers a direct route to detecting coherent phonon transport in real materials.

To test this hypothesis, we investigate thermal transport in CuCl. CuCl crystallizes in the zinc blende structure~\cite{CuClStructure}, an uncommon structure for a metal halide, which more typically forms in rocksalt or cesium chloride phases. This structural anomaly is accompanied by pronounced anharmonic behavior: CuCl exhibits negative thermal expansion~\cite{CuClTE}, non-Lorentzian phonon spectral functions observed in neutron and Raman scattering~\cite{CuClNeutronTO,CuClRaman1,CuClRaman2,CuClRaman3}, and multiple pressure-induced phase transitions~\cite{CuClPT1,CuClPT2}.

Most relevant to the present study, CuCl possesses an unexpectedly low lattice thermal conductivity~\cite{CuCl_kappa}. Moreover, its thermal conductivity decreases with increasing pressure, contrary to typical behavior. This puzzling trend has been attributed to enhanced scattering phase space for longitudinal acoustic phonons under pressure~\cite{CuClkappa_theory1}. These properties—low $\kappa$, strong anharmonicity, and spectral broadening—make CuCl an ideal candidate for observing signatures of coherent thermal transport.

\begin{figure}
    \centering
    \includegraphics[width=0.99\linewidth]{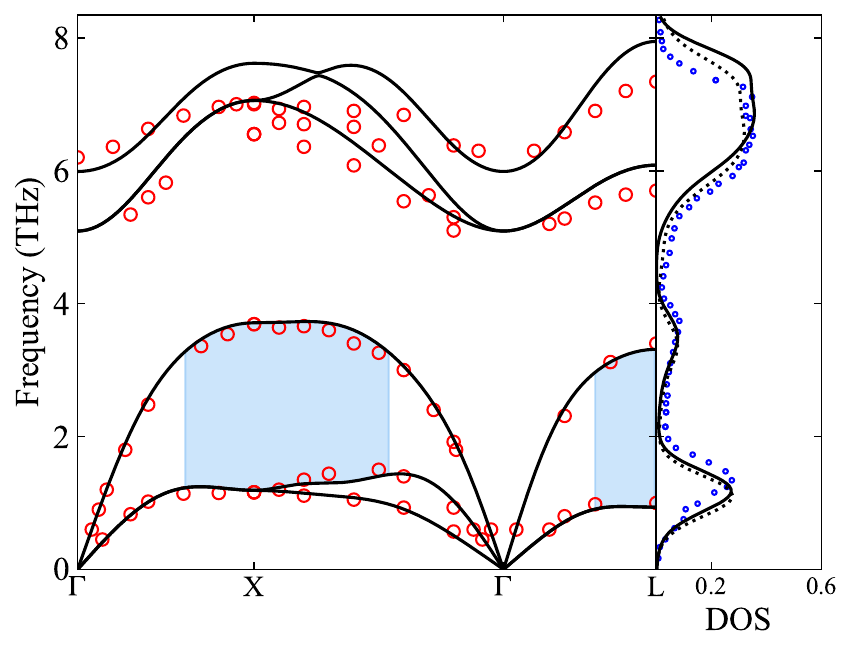}
    \caption{Calculated and measured phonon band structure of CuCl. The full black lines are SSCHA calculations, while experiments are shown in points (red~\cite{CuCl_phonons_exp} and blue~\cite{CuClkappa_theory1}). Dashed lines represent phonon density of states calculated using phonon spectral functions. Shaded regions represent phonon frequency nesting responsible for spectral signatures in Fig.~\ref{fig:ls}.}
    \label{fig:phonons}
\end{figure}

Given the strong anharmonicity of CuCl, we compute its vibrational properties using the stochastic self-consistent harmonic approximation (SSCHA)~\cite{SSCHA1,SSCHA2,SSCHA4,SSCHA3,SSCHA5}. (For computational details, see the Supplementary Material~\cite{SSCHA1,SSCHA2,SSCHA4,mace,QE1,QE2,QE3,PBESOL,pseudos}.) Figure~\ref{fig:phonons} shows the calculated phonon band structure at 0 GPa and 0 K, in comparison with experimental data from inelastic neutron scattering~\cite{CuCl_phonons_exp}. The agreement is excellent for the heat-carrying acoustic phonons, which are most relevant for thermal transport. While the match for optical phonons is less accurate, the deviations remain within acceptable bounds. The discrepancy for optical modes is also reflected in the phonon density of states (shown in the side panel of Fig.~\ref{fig:phonons}), where experimental data exhibit greater softening of the optical modes. Inclusion of dynamical effects on the phonon quasiparticles (dashed line in the figure) improves the agreement, capturing part of the observed softening. The calculated phonon density of states is weighted by the neutron scattering cross section of each atomic species and convoluted with a Gaussian to account for experimental energy resolution.

\begin{figure}
    \centering
    \includegraphics[width=0.99\linewidth]{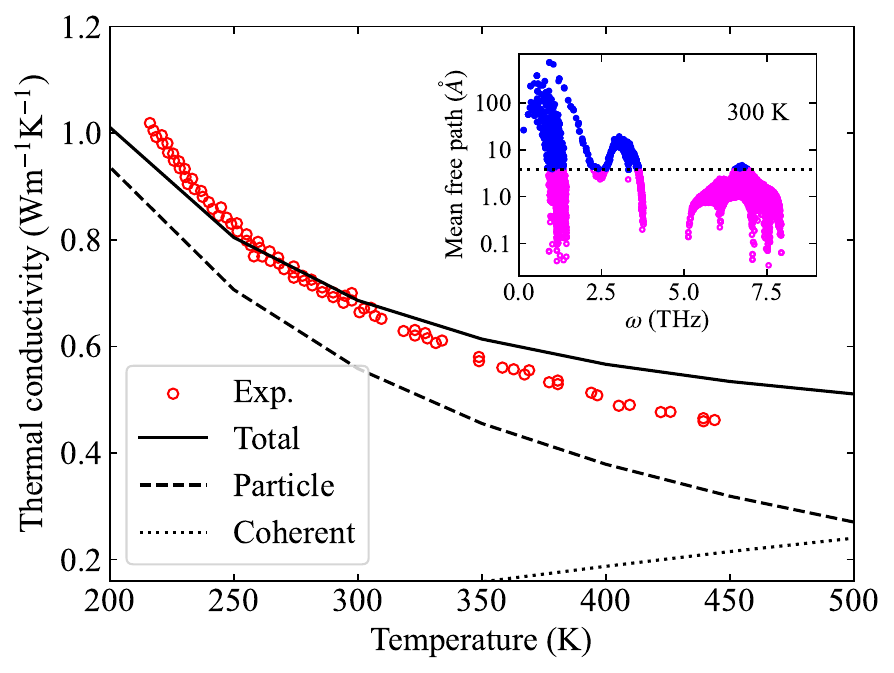}
    \caption{Calculated and measured lattice thermal conductivity $\kappa$ of CuCl. The full black line is total $\kappa$, dashed line is the diagonal contribution and pointed line is the off-diagonal, coherent, contribution. Experiment~\cite{CuCl_kappa} is shown in red points. The inset shows the phonon mean free path  at 300 K with dashed line being the lattice constant indicating boundary for diffusive transport.}
    \label{fig:kappa}
\end{figure}

We compute the lattice thermal conductivity of CuCl in the perturbative limit across a range of temperatures (see Fig.~\ref{fig:kappa}). The agreement with experimental measurements~\cite{CuCl_kappa} is very good, particularly given that earlier first-principles calculations significantly overestimated $\kappa$ in this material~\cite{CuClkappa_theory1, CuCl_kappa_theory2}. This improvement is largely due to the inclusion of anharmonicity-renormalized third-order force constants in our SSCHA calculations. Nevertheless, we note that the agreement is likely partially fortuitous and may result from a cancellation of errors (see Supplementary Material for discussion).

The temperature dependence of the residual discrepancy between our calculation and experiment points to fourth-order anharmonicity as the main source of the mismatch, since it is the only mechanism that is strongly temperature dependent. Scattering from lattice imperfections would instead be temperature independent, and a previous study~\cite{CuClkappa_theory1} estimated that grain sizes on the order of only a few tens of nanometers would be required to bring third-order-only calculations into agreement with experiment—sizes not supported by any experimental observations. Finally, superionic behaviour in CuCl has been reported~\cite{BOYCE1980237} only at temperatures above 550 K, roughly 100 K higher than the maximum temperature considered in the thermal conductivity measurements, and therefore cannot account for the observed temperature-dependent discrepancy.

Despite CuCl's structural simplicity—only two atoms per primitive unit cell—the coherent contribution to its lattice thermal conductivity is non-negligible, particularly at elevated temperatures. Our results show that this coherent component strongly renormalizes the temperature dependence of $\kappa$, producing an almost flat thermal conductivity curve at high temperatures. This behaviour is not observed experimentally, reinforcing the conclusion that higher-order anharmonic effects, especially quartic (fourth-order) interactions, are essential for accurately capturing thermal transport in CuCl at elevated temperatures~\cite{cucl_4th_order, newcucl}.


The inset of Fig.~\ref{fig:kappa} shows the phonon mean free paths calculated at 300 K. The results indicate that heat transport is dominated by low-frequency acoustic phonons, particularly the longitudinal modes. The dashed line denotes the lattice constant, which serves as an approximate lower bound for diffusive (particle-like) transport. Phonons with mean free paths shorter than this scale are unlikely to contribute significantly via conventional mechanism. At 300 K, nearly all optical phonons fall into this sub-diffusive regime. In addition, a substantial fraction of acoustic modes also exhibit short mean free paths. This is primarily due to their low group velocities—especially for phonons away from the Brillouin zone center, as seen in Fig.~\ref{fig:phonons}. 

\begin{figure}
    \centering
    \includegraphics[width=0.99\linewidth]{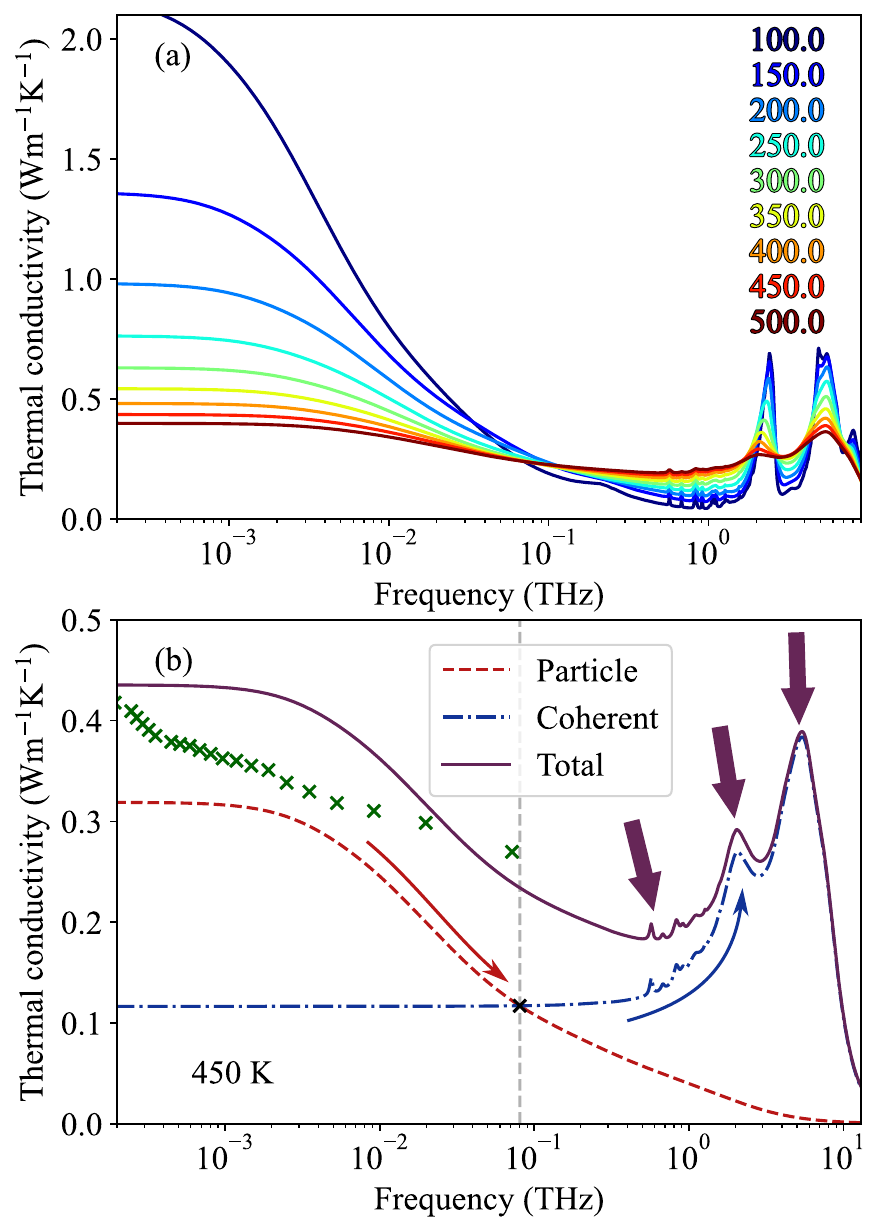}
    \caption{Dynamical lattice thermal conductivity. (a) Total dynamical lattice thermal conductivity at different temperatures. (b) Dynamical lattice thermal conductivity at 450 K. The black cross denotes the frequency at which coherent transport gives dominant contribution. The green crosses are results for the mean free path cumulative lattice thermal conductivity expressed in appropriate frequency.}
    \label{fig:ac}
\end{figure}

We now present the dynamical lattice thermal conductivity $\kappa(\nu)$, calculated in the perturbative regime. Figure~\ref{fig:ac}(a) shows $\kappa(\nu)$ at several temperatures. At low temperatures, the thermal conductivity exhibits a rapid decay with increasing frequency, consistent with Eq.~\ref{eq:pert_diag}, due to the longer phonon lifetimes. As temperature increases, the decay becomes progressively weaker, and at very high temperatures, $\kappa(\nu)$ flattens, resembling the nearly frequency-independent behavior observed in amorphous materials. Superimposed on this background, we observe distinct peaks in $\kappa(\nu)$ at higher frequencies, as predicted by Eq.~\ref{coherent_dyn_kappa}. These features are most pronounced at low temperatures, where narrow phonon linewidths allow coherent contributions to emerge clearly. At elevated temperatures, increased anharmonic broadening smears out these features, resulting in a smoother and more uniform frequency response. Notably, in certain frequency windows at high temperature, $\kappa(\nu)$ exceeds the static ($\nu \to 0$) value.

In Fig.\ref{fig:ac}(b), we show the decomposition of the dynamical lattice thermal conductivity at 450 K into particle-like and coherent contributions. As expected from Eq.\ref{eq:pert_diag}, the particle-like component decays monotonically with increasing frequency. In contrast, the coherent contribution exhibits a non-monotonic response with distinct spectral features at higher frequencies, consistent with Eq.~\ref{coherent_dyn_kappa}. At this temperature, the coherent component begins to dominate above 0.1 THz, indicating that even low-frequency transport is significantly influenced by mode interference effects. Beyond 1 THz, the transport is almost entirely coherent in nature. Notably, pronounced peaks appear around 2 THz and 6 THz, with amplitudes comparable to the static thermal conductivity. These features provide a clear spectral fingerprint of coherent phonon transport and represent a direct, testable prediction for future experiments.

We now turn to the interpretation of time-domain thermoreflectance (TDTR)~\cite{TDTR, aSi_TDTR,GangChen_TDTR,MX2_TDTR} and frequency-domain thermoreflectance (FDTR)~\cite{FDTR, FDTR2, FDTR3} experiments, which effectively probe the frequency-dependent lattice thermal conductivity $\kappa(\nu)$. It is worth noting that in these types of experiments the heating is applied at a particular spot on the sample (more precisely, on a thin metallic transducer layer deposited on top of it), and thus our assumption of uniform heating (Eq.~\ref{eq:fourier_law}) does not strictly hold. However, we believe that for a sufficiently large heating spot the spatial dependence of the lattice thermal conductivity can be neglected, as is typically done in current experimental analyses. In these techniques, the decay of $\kappa(\nu)$ with frequency is commonly attributed to the finite thermal penetration depth of the modulated heat source~\cite{Cahill_kappa_nu, Mingo_TDTR}, which is given by: $L_p = \sqrt{\frac{\kappa (\omega_{LP})}{\pi C \omega _{LP}}}$, where $C$ is the volumetric heat capacity and $\omega_{LP}$ is the modulation frequency. Within this framework, phonons with mean free paths $l_{\mathbf{q},j} = |\mathbf{v}_{\mathbf{q},j}|\tau_{\mathbf{q},j}$ longer than $L_p$ are assumed to travel ballistically and, therefore, do not contribute to the measured conductivity. This motivates the definition of a cumulative thermal conductivity:
\begin{align*}
    \kappa ^{x,y} (L_p) = \frac{1}{NV}\sum _{\mathbf{q},j} \kappa ^{x,y} _{\mathbf{q},j}\Theta(L_p - l_{\mathbf{q},j}), \numberthis
    \label{cumm_kappa}
\end{align*}
where $\Theta(x)$ is the Heaviside function. By associating $\omega_{LP}$ with a driving frequency $\nu$, this model allows comparison between measured $\kappa(\nu)$ and mean free path-limited predictions.

In the Supplementary Material, we show that this approximation holds well in systems where particle-like transport dominates (high lattice thermal conductivity materials). The agreement improves further when we assume that phonon mean free paths are bounded by the penetration depth—that is, modes with mean free paths exceeding the penetration depth scatter at the effective boundary defined by it. However, in our case (see Fig.~\ref{fig:ac}(b), red crosses), the cumulative model based on Eq.~\ref{cumm_kappa} does not agree with the directly computed $\kappa(\nu)$. This discrepancy confirms that, in materials where coherent or strongly anharmonic transport mechanisms are significant—such as CuCl, amorphous solids, or alloys—the thermal penetration depth and phonon mean free paths do not longer directly correlate with lattice thermal conductivity. As a result, traditional TDTR/FDTR interpretations based solely on particle-like transport may not fully capture the underlying physics in such systems.
 
\begin{figure}
    \centering
    \includegraphics[width=0.99\linewidth]{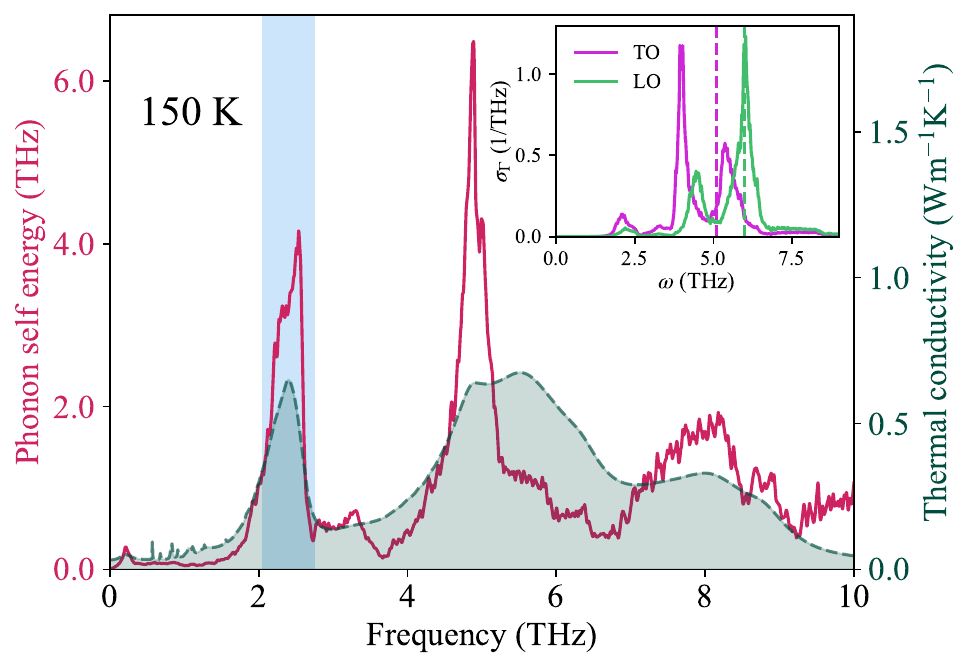}
    \caption{Phonon self-energy (just the imaginary part) of optical phonon modes at 150 K in CuCl. The gray background is the coherent contribution to dynamical lattice thermal conductivity at 150 K. The inset show phonon spectral functions of transverse and longitudinal optical modes.}
    \label{fig:ls}
\end{figure}

Finally, we analyze the microscopic origin of the peaks observed in $\kappa(\nu)$. As indicated by Eq.~\ref{coherent_dyn_kappa}, these features arise from the nesting of phonon frequencies—that is, pairs of modes with the same wave vector and closely spaced frequencies. Mathematically, the coherent contribution resembles a two-phonon joint density of states, broadened by the sum of phonon linewidths and weighted by the generalized group velocities. As such, it bears a close connection to the phonon self-energy, specifically for zone-center ($\Gamma$-point) modes. This connection is illustrated in Fig.~\ref{fig:ls}, where we compare the calculated phonon self-energy with the coherent component of $\kappa(\nu)$. Both quantities exhibit a pronounced peak near 2 THz, which originates from strong coupling between longitudinal acoustic (LA) and transverse acoustic (TA) modes. This mode nesting (highlighted in Fig.~\ref{fig:phonons}) is a primary source of anharmonicity in CuCl. Additionally, the phonon self-energy displays a second strong peak around 5 THz, associated with coupling between long-wavelength transverse optical (TO) and acoustic phonons. This feature lies near the auxiliary phonon frequency and reflects another important channel of anharmonic interaction. 

The inset of Fig.\ref{fig:ls} shows the phonon spectral functions of the $\Gamma$-point optical modes at 150 K. Our calculations reveal significant splitting into satellite peaks, in agreement with Raman spectroscopy experiments\cite{CuClRaman1, CuClRaman2, CuClRaman3}. In addition to the pronounced high-frequency satellite structure associated with the transverse and longitudinal optical (TO and LO) modes, we also observe a smaller satellite peak around 2 THz. As we have noted in the case of phonon self-energy, this lower-frequency feature originates from nesting between longitudinal and transverse acoustic (LA and TA) phonon bands.

\section{Discussion}
We have proposed frequency-resolved lattice thermal conductivity $\kappa(\nu)$ as a direct probe of coherent phonon transport. Specifically, we have shown that peaks in $\kappa(\nu)$ arise from the nesting of phonon modes with the same wave vector—an unambiguous signature of coherence. However, several challenges complicate the experimental detection of this effect. Current time-domain (TDTR) and frequency-domain (FDTR) thermoreflectance techniques typically operate at modulation frequencies up to $\sim$100 MHz, whereas coherent features in $\kappa(\nu)$ are expected to appear at THz frequencies, comparable to the phonon frequency size. This represents a discrepancy of four orders of magnitude between current experimental capabilities and the frequency range where coherent signatures are most pronounced.

Moreover, isolating coherent contributions requires that particle-like transport be sufficiently suppressed. While particle transport decays quickly with frequency when phonon lifetimes are long, in such cases the coherent contribution tends to be small. Conversely, in strongly anharmonic systems where coherent effects are expected to be large, phonon lifetimes are short, and the particle component decays more slowly—masking coherent features, as illustrated in Fig.~\ref{fig:ac}.

These limitations suggest that amorphous materials or alloys may offer a more promising platform for detecting coherence. Indeed, TDTR measurements on amorphous Si and SiO$_2$ show broad, plateau-like behavior in $\kappa(\nu)$ at high temperatures—similar to our results for CuCl. However, such plateaus may also result from short phonon mean free paths and do not provide conclusive evidence of coherent transport. In contrast, an increase in $\kappa(\nu)$ with frequency constitutes a more definitive signature, since the particle-like contribution is inherently monotonic and decreasing. While such behavior has been reported for amorphous materials~\cite{Regner2013}, the experimental uncertainties in those studies remain too large for a conclusive interpretation.

We note that, in principle, even particle-like transport could produce a weak peak in $\kappa(\nu)$ if the phonon spectral function $\sigma_{\mathbf{q},j}(\Omega)$ features satellite peaks. In such a case, a resonant frequency could arise from the difference between the main and satellite peaks. However, for this effect to significantly influence $\kappa(\nu)$, many phonon modes would need to exhibit satellites at the same frequency offset—an improbable scenario. Thus, we conclude that any observed increase in $\kappa(\nu)$ with frequency is a strong and likely unique signature of coherent phonon transport.

Furthermore, these types of experiments could, in principle, be used to directly probe topologically protected phonon surface states. To date, there has been no unambiguous experimental observation of such topological phonon modes. These surface states are highly localized (i.e., they are nearly flat and possess vanishing group velocity) and therefore do not contribute to the static thermal conductivity $\kappa$. However, precisely because of their localized nature, they can give rise to distinct peaks in the dynamical lattice thermal conductivity $\kappa(\nu)$. Depending on the surface termination of a given topological material, these modes may or may not exist, and thus their presence—or absence—would be directly reflected in the measured $\kappa(\nu)$. Moreover, since these states are confined to the surface, they are difficult to detect using conventional scattering techniques. In contrast, TDTR and FDTR measurements are inherently surface-sensitive (especially in the high frequency regimes where heat penetration depth is very small), which makes them ideally suited for the experimental verification of topological phonon surface modes.

Finally, it is important to emphasize the potential role of dynamical lattice thermal conductivity in the design of next-generation electronic devices. A large portion of modern electronics operates in the GHz regime, which, according to Eq.~\ref{eq:pert_diag}, corresponds to frequencies where the effective thermal conductivity of silicon may experience a significant drop. At these and higher frequencies, coherent phonon contributions become non-negligible, and it is therefore crucial to account for them when developing strategies for thermal management in future devices. Promising avenues for further research include the study of interface thermal conductivity at finite frequencies and the engineering of dynamical $\kappa$ through alloying, superlattices, and other nanostructured materials.

\section{Conclusion}
In conclusion, we have developed a framework to identify signatures of coherent phonon transport through the frequency-dependent lattice thermal conductivity, $\kappa(\nu)$. Unlike particle-like transport, which decays monotonically with frequency, the coherent contribution exhibits distinct spectral peaks—arising from the nesting of phonon modes with the same wave vector. These effects are expected to exist universally across all classes of materials, but they should be most pronounced in strongly anharmonic crystals and in systems with broken discrete translational symmetry. Applying this formalism to the highly anharmonic material CuCl, we uncover clear non-monotonic features in $\kappa(\nu)$ associated with acoustic phonon mode coupling. These peaks offer a direct spectral fingerprint of coherence in heat transport. Finally, we discuss the challenges and opportunities for experimental detection, highlighting the limitations of current thermoreflectance techniques and the promise of extending such measurements to disordered or amorphous systems. Our work provides a pathway for directly probing and validating coherent thermal transport in complex materials.

\section{Acknowledgments}

The author wishes to acknowledge Ion Errea for inspiring discussions and constant encouragement. This work was supported by the European Research Council (ERC) under the European Union's Horizon 2020 research and innovation program (grant agreement No. 802533). 

\bibliographystyle{unsrt}
\bibliography{main}

	\onecolumngrid
    \renewcommand{\figurename}{Supplementary Figure}
    \renewcommand{\tablename}{Supplementary Table}
    \setcounter{figure}{0}
	\newpage
	\newpage
	\newpage

\section{Supplementary material: \\ Signatures of coherent phonon transport in frequency dependent lattice thermal conductivity}

\subsection{Computational details}

The structure of CuCl at 0 K was obtained by minimizing the total free energy using the SSCHA~\cite{SSCHA1,SSCHA2,SSCHA4}. The second-order force constants were also obtained from this minimization. Third-order force constants were subsequently calculated for this structure using the SSCHA code with 20,000 random configurations. The energies, stresses, and forces required for the SSCHA calculations were generated using the MACE interatomic potential~\cite{mace}.

The MACE potential was fitted as follows. First, harmonic DFPT dynamical matrices for CuCl were calculated on a $4\times 4\times 4$ $\mathbf{q}$-point grid. SSCHA was then used to generate random structures over a temperature range from 0 to 550 K for a $3\times 3\times 3$ supercell (54 atoms). Self-consistent field calculations were performed using Quantum ESPRESSO~\cite{QE1,QE2,QE3} with the PBESOL exchange-correlation functional~\cite{PBESOL} and ultrasoft pseudopotentials~\cite{pseudos} to represent the ions. The electronic wavefunctions were expanded in a plane-wave basis with a cutoff of 80 Ry, and the supercell Brillouin zone was sampled with a $3\times 3\times 3$ $\mathbf{k}$-point grid.

A total of 808 configurations were used for the training and validation sets, and 7,842 configurations were used for testing. The mean-square errors for the test set are 1.1 meV/atom for energies, 5.6 meV/$\mathring{A}$ for forces, and 0.4 meV/$\mathring{A} ^3$ for stresses.

Finally, thermal conductivity calculations were performed using SSCHA code~\cite{GK-main}. Phonon modes were sampled on $28\times 28\times 28$ grid. The energy conservation for phonon-phonon scattering was accounted by Gaussian smearing with an adaptive smearing parameter. In the calculation of the phonon lifetimes and lineshapes only third-order anharmonicity was used through the bubble diagram. 

\newpage

\subsection{Perturbative limit for dynamical lattice thermal conductivity}

We start from the full equation for the lattice thermal conductivity: \\
\begin{align*}
    \kappa ^{xy} (\nu) = &\frac{\pi\beta ^2k_{B}}{NV}\sum _{\mathbf{q},j,j'}v^{x}_{\mathbf{q},j,j'}v^{y*}_{\mathbf{q},j,j'}\omega _{\mathbf{q},j}\omega _{\mathbf{q},j'}\frac{e^{\beta\nu} - 1}{\beta\nu}\times \\ 
    &\int_{-\infty}^{\infty}\textrm{d}\Omega (1+\frac{\Omega}{\Omega + \nu})\frac{e^{\beta\Omega}}{\left(e^{\beta\Omega} - 1\right)\left(e^{\beta(\Omega + \nu)} - 1\right)}\sigma _{\mathbf{q},j'}(\Omega)\sigma_{\mathbf{q},j}(\Omega + \nu). \numberthis
    \label{eq:dyn_kappa_init}
\end{align*}

We assume a Lorentzian shape of the spectral function:

\begin{align*}
    \sigma _{\mathbf{q},j} (\Omega) = \frac{1}{2\pi}\left(\frac{\Gamma _{\mathbf{q},j}}{(\Omega - \omega _{\mathbf{q},j})^2 + \Gamma ^2 _{\mathbf{q},j}} + \frac{\Gamma _{\mathbf{q},j}}{(\Omega + \omega _{\mathbf{q},j})^2 + \Gamma ^2 _{\mathbf{q},j}}\right). \numberthis
    \label{LA_spec_func} 
\end{align*}

The difficult part is to estimate the value of the integral. We will do it by using the expression for the convolution of two Lorentzians ($L_1(\Omega) = \frac{1}{\pi}\frac{\Gamma _1}{\Omega ^2 + \Gamma _1^2}$):
\begin{align*}
    L(\nu) = L_1(\Omega)\star L_2(\Omega) = \int_{-\infty}^{\infty}\textrm{d}\Omega \frac{1}{\pi ^2}\frac{\Gamma _1}{(\Omega - \nu) ^2 + \Gamma _1^2}\frac{\Gamma _2}{\Omega ^2 + \Gamma _2^2} = \frac{1}{\pi}\frac{\Gamma _1 + \Gamma _2}{\nu ^2 + (\Gamma _1 + \Gamma _2)^2}. \numberthis
    \label{conv_la}
\end{align*}

First let's analyse the case for the particle contribution to the lattice thermal conductivity (diagonal $j=j'$ case). Multiplying Lorentzians inside the integral~\ref{eq:dyn_kappa_init} we get four distinct contributions:
\begin{align*}
    &\int_{-\infty}^{\infty}\textrm{d}\Omega (1+\frac{\Omega}{\Omega + \nu})\frac{e^{\beta\Omega}}{\left(e^{\beta\Omega} - 1\right)\left(e^{\beta(\Omega + \nu)} - 1\right)}\sigma _{\mathbf{q},j'}(\Omega)\sigma_{\mathbf{q},j}(\Omega + \nu) = \\
    &\int_{-\infty}^{\infty}\textrm{d}\Omega (1+\frac{\Omega}{\Omega + \nu})\frac{e^{\beta\Omega}}{\left(e^{\beta\Omega} - 1\right)\left(e^{\beta(\Omega + \nu)} - 1\right)}\frac{1}{4\pi^2}\frac{\Gamma _{\mathbf{q},j}}{(\Omega - \omega _{\mathbf{q},j})^2 + \Gamma ^2 _{\mathbf{q},j}}\frac{\Gamma _{\mathbf{q},j}}{(\Omega +\nu - \omega _{\mathbf{q},j})^2 + \Gamma ^2 _{\mathbf{q},j}} + \\
    &\int_{-\infty}^{\infty}\textrm{d}\Omega (1+\frac{\Omega}{\Omega + \nu})\frac{e^{\beta\Omega}}{\left(e^{\beta\Omega} - 1\right)\left(e^{\beta(\Omega + \nu)} - 1\right)}\frac{1}{4\pi^2}\frac{\Gamma _{\mathbf{q},j}}{(\Omega - \omega _{\mathbf{q},j})^2 + \Gamma ^2 _{\mathbf{q},j}}\frac{\Gamma _{\mathbf{q},j}}{(\Omega +\nu + \omega _{\mathbf{q},j})^2 + \Gamma ^2 _{\mathbf{q},j}} + \\
    &\int_{-\infty}^{\infty}\textrm{d}\Omega (1+\frac{\Omega}{\Omega + \nu})\frac{e^{\beta\Omega}}{\left(e^{\beta\Omega} - 1\right)\left(e^{\beta(\Omega + \nu)} - 1\right)}\frac{1}{4\pi^2}\frac{\Gamma _{\mathbf{q},j}}{(\Omega + \omega _{\mathbf{q},j})^2 + \Gamma ^2 _{\mathbf{q},j}}\frac{\Gamma _{\mathbf{q},j}}{(\Omega +\nu - \omega _{\mathbf{q},j})^2 + \Gamma ^2 _{\mathbf{q},j}} + \\
    &\int_{-\infty}^{\infty}\textrm{d}\Omega (1+\frac{\Omega}{\Omega + \nu})\frac{e^{\beta\Omega}}{\left(e^{\beta\Omega} - 1\right)\left(e^{\beta(\Omega + \nu)} - 1\right)}\frac{1}{4\pi^2}\frac{\Gamma _{\mathbf{q},j}}{(\Omega + \omega _{\mathbf{q},j})^2 + \Gamma ^2 _{\mathbf{q},j}}\frac{\Gamma _{\mathbf{q},j}}{(\Omega +\nu + \omega _{\mathbf{q},j})^2 + \Gamma ^2 _{\mathbf{q},j}}. 
\end{align*}

Because of the additional terms in the integral we can not use the identity for the convolution of Lorentzians. For that reason, we will estimate values of these additional terms at frequencies $\Omega$ where these Lorentzians peak ($\Omega = \omega _{\mathbf{q},j}$). For example the first term becomes:
\begin{align*}
    &\int_{-\infty}^{\infty}\textrm{d}\Omega (1+\frac{\Omega}{\Omega + \nu})\frac{e^{\beta\Omega}}{\left(e^{\beta\Omega} - 1\right)\left(e^{\beta(\Omega + \nu)} - 1\right)}\frac{1}{4\pi^2}\frac{\Gamma _{\mathbf{q},j}}{(\Omega - \omega _{\mathbf{q},j})^2 + \Gamma ^2 _{\mathbf{q},j}}\frac{\Gamma _{\mathbf{q},j}}{(\Omega +\nu - \omega _{\mathbf{q},j})^2 + \Gamma ^2 _{\mathbf{q},j}} \approx \\
    &\frac{1}{4\pi^2} (1+\frac{\omega _{\mathbf{q},j}}{\omega _{\mathbf{q},j} + \nu})\frac{e^{\beta\omega _{\mathbf{q},j}}}{\left(e^{\beta\omega _{\mathbf{q},j}} - 1\right)\left(e^{\beta(\omega _{\mathbf{q},j} + \nu)} - 1\right)}\int_{-\infty}^{\infty}\textrm{d}\Omega\frac{\Gamma _{\mathbf{q},j}}{(\Omega - \omega _{\mathbf{q},j})^2 + \Gamma ^2 _{\mathbf{q},j}}\frac{\Gamma _{\mathbf{q},j}}{(\Omega +\nu - \omega _{\mathbf{q},j})^2 + \Gamma ^2 _{\mathbf{q},j}} = \\
    &\frac{1}{4\pi} (1+\frac{\omega _{\mathbf{q},j}}{\omega _{\mathbf{q},j} + \nu})\frac{e^{\beta\omega _{\mathbf{q},j}}}{\left(e^{\beta\omega _{\mathbf{q},j}} - 1\right)\left(e^{\beta(\omega _{\mathbf{q},j} + \nu)} - 1\right)}\frac{2\Gamma _{\mathbf{q},j}}{\nu^2 + 4\Gamma ^2 _{\mathbf{q},j}} \approx \\
    & \frac{1}{2\pi}\frac{e^{\beta\omega _{\mathbf{q},j}}}{\left(e^{\beta\omega _{\mathbf{q},j}} - 1\right)\left(e^{\beta\omega _{\mathbf{q},j}} - 1\right)}\frac{2\Gamma _{\mathbf{q},j}}{\nu^2 + 4\Gamma ^2 _{\mathbf{q},j}} = \frac{1}{2\pi}\frac{e^{\beta\omega _{\mathbf{q},j}}}{\left(e^{\beta\omega _{\mathbf{q},j}} - 1\right)\left(e^{\beta\omega _{\mathbf{q},j}} - 1\right)}\frac{\tau _{\mathbf{q},j}}{\nu^2\tau _{\mathbf{q},j}^2  + 1}.
\end{align*}

In the final line we assumed that the fastest changing part with $\nu$ comes from the final Lorentzian, while exponential part is slowly varying with $\nu$. In that case we can approximate $\nu = 0$ outside of the Lorentzian since that is where Lorentzian peaks. Similar reasoning can be applied to the rest of terms which finally yields:
\begin{align*}
    \kappa ^{xy} (\nu) = \frac{1}{NV}\sum_{\mathbf{q},j} \left( \frac{\kappa ^{xy}_{\mathbf{q},j}}{\nu^2\tau _{\mathbf{q},j}^2  + 1} + \frac{\kappa ^{xy}_{\mathbf{q},j}}{(\nu + 2\omega_{\mathbf{q},j})^2\tau _{\mathbf{q},j}^2  + 1}\right). \numberthis
    \label{diag_pert}
\end{align*}
The second terms is much smaller obviously and can be safely discarded. In the limit $\nu\rightarrow 0$ this reduces to the solution of the Boltzmann transport equation in the relaxation time approximation as we would expect.

The derivation of the coherent contribution $j\neq j'$ is somewhat more complicated. We again have four different terms coming from multiplication of different Lorentzians. Let's focus on the first product of Lorentzians (for the rest similar approach applies):
\begin{align*}
    I_1 =& \frac{1}{4\pi^2}\int _{-\infty}^{\infty}\textrm{d}\Omega (1+\frac{\Omega}{\Omega + \nu})\frac{e^{\beta\Omega}}{\left(e^{\beta\Omega} - 1\right)\left(e^{\beta(\Omega + \nu)} - 1\right)}\frac{\Gamma _{\mathbf{q},j'}}{(\Omega - \omega _{\mathbf{q},j'})^2 + \Gamma _{\mathbf{q},j'}}\frac{\Gamma _{\mathbf{q},j}}{(\Omega + \nu- \omega _{\mathbf{q},j})^2 + \Gamma _{\mathbf{q},j}} = \\
    &\{\text{Substitution: } \Omega' = \Omega - \omega _{\mathbf{q},j'}; \Omega = \Omega ' + \omega _{\mathbf{q},j'}; \textrm{d}\Omega ' = \textrm{d}\Omega\} = \numberthis \label{step_1} \\
    &\frac{1}{4\pi^2}\int _{-\infty}^{\infty}\textrm{d}\Omega' (1+\frac{\Omega' + \omega _{\mathbf{q},j'}}{\Omega' + \omega _{\mathbf{q},j'} + \nu})\frac{e^{\beta(\Omega' + \omega _{\mathbf{q},j'})}}{\left(e^{\beta(\Omega' + \omega _{\mathbf{q},j'})} - 1\right)\left(e^{\beta(\Omega' + \omega _{\mathbf{q},j'} + \nu)} - 1\right)}\frac{\Gamma _{\mathbf{q},j'}}{\Omega '^2 + \Gamma _{\mathbf{q},j'}}\frac{\Gamma _{\mathbf{q},j}}{(\Omega' + \omega _{\mathbf{q},j'} + \nu- \omega _{\mathbf{q},j})^2 + \Gamma _{\mathbf{q},j}}. 
\end{align*}
We then take the values of the non-Lorentzian terms inside the integral at $\Omega' = 0$ (we would get same result if we approximated them at the value where second Lorentzian peaks):
\begin{align*}
    I_1 = \frac{1}{4\pi}(1+\frac{\omega _{\mathbf{q},j'}}{\omega _{\mathbf{q},j'} + \nu})\frac{e^{\beta\omega _{\mathbf{q},j'}}}{\left(e^{\beta\omega _{\mathbf{q},j'}} - 1\right)\left(e^{\beta(\omega _{\mathbf{q},j'} + \nu)} - 1\right)}\frac{\Gamma _{\mathbf{q},j'} + \Gamma _{\mathbf{q},j}}{(-\nu + (\omega _{\mathbf{q},j} - \omega _{\mathbf{q},j'}))^2 + (\Gamma _{\mathbf{q},j'} + \Gamma _{\mathbf{q},j})^2}. \numberthis
    \label{step_2}
\end{align*}
Finally, we will substitute $\nu$ with a value where the Lorentzian peaks everywhere except in the Lorentzian to get:
\begin{align*}
    I_1 &= \frac{1}{4\pi}(1+\frac{\omega _{\mathbf{q},j'}}{\omega _{\mathbf{q},j}})\frac{e^{\beta\omega _{\mathbf{q},j'}}}{\left(e^{\beta\omega _{\mathbf{q},j'}} - 1\right)\left(e^{\beta\omega _{\mathbf{q},j}} - 1\right)}\frac{\Gamma _{\mathbf{q},j'} + \Gamma _{\mathbf{q},j}}{(-\nu + (\omega _{\mathbf{q},j} - \omega _{\mathbf{q},j'}))^2 + (\Gamma _{\mathbf{q},j'} + \Gamma _{\mathbf{q},j})^2}. \numberthis
    \label{step_3}
\end{align*}
Other three terms are:
\begin{align*}
    I_2 &= \frac{1}{4\pi}(1-\frac{\omega _{\mathbf{q},j'}}{\omega _{\mathbf{q},j}})\frac{e^{\beta\omega _{\mathbf{q},j'}}}{\left(e^{\beta\omega _{\mathbf{q},j'}} - 1\right)\left(e^{-\beta\omega _{\mathbf{q},j}} - 1\right)}\frac{\Gamma _{\mathbf{q},j'} + \Gamma _{\mathbf{q},j}}{(\nu + (\omega _{\mathbf{q},j} + \omega _{\mathbf{q},j'}))^2 +(\Gamma _{\mathbf{q},j'} + \Gamma _{\mathbf{q},j})^2 } \\
    I_3 &= \frac{1}{4\pi}(1-\frac{\omega _{\mathbf{q},j'}}{\omega _{\mathbf{q},j}})\frac{e^{-\beta\omega _{\mathbf{q},j'}}}{\left(e^{-\beta\omega _{\mathbf{q},j'}} - 1\right)\left(e^{\beta\omega _{\mathbf{q},j}} - 1\right)}\frac{\Gamma _{\mathbf{q},j'} + \Gamma _{\mathbf{q},j}}{(\nu - (\omega _{\mathbf{q},j} + \omega _{\mathbf{q},j'}))^2 +(\Gamma _{\mathbf{q},j'} + \Gamma _{\mathbf{q},j})^2 } \\
    I_4 &= \frac{1}{4\pi}(1+\frac{\omega _{\mathbf{q},j'}}{\omega _{\mathbf{q},j}})\frac{e^{-\beta\omega _{\mathbf{q},j'}}}{\left(e^{-\beta\omega _{\mathbf{q},j'}} - 1\right)\left(e^{-\beta\omega _{\mathbf{q},j}} - 1\right)}\frac{\Gamma _{\mathbf{q},j'} + \Gamma _{\mathbf{q},j}}{(\nu + (\omega _{\mathbf{q},j} - \omega _{\mathbf{q},j'}))^2 +(\Gamma _{\mathbf{q},j'} + \Gamma _{\mathbf{q},j})^2 }
\end{align*}
Second and third term contribute most at $\nu = \pm (\omega _{\mathbf{q},j} + \omega _{\mathbf{q},j'})$, which will give a relatively small contribution at low and medium frequencies for which we are mostly interested. Keeping in mind that the sum runs over $jj'$ and $j'j$ this can be further simplified to the equation reported in the main part:
\begin{align*}
        \kappa ^{x,y} (\nu) =  &\frac{\beta ^2k_{B}}{4NV}\sum _{\mathbf{q},j>j'}v^{x}_{\mathbf{q},j,j'}v^{y*}_{\mathbf{q},j,j'}\frac{e^{\beta(\omega _{\mathbf{q},j} - \omega _{\mathbf{q},j'})} - 1}{\beta(\omega _{\mathbf{q},j} - \omega _{\mathbf{q},j'})}\frac{e^{\beta\omega _{\mathbf{q},j'}}(\omega _{\mathbf{q},j} + \omega _{\mathbf{q},j'})^2}{(e^{\beta\omega _{\mathbf{q},j}} - 1)(e^{\beta\omega _{\mathbf{q},j'}} - 1)}\times \\ &\left(\frac{\Gamma _{\mathbf{q},j} + \Gamma _{\mathbf{q},j'}}{(\nu + \omega _{\mathbf{q},j} - \omega_{\mathbf{q},j'})^2+(\Gamma _{\mathbf{q},j} + \Gamma _{\mathbf{q},j'})^2} + \frac{
    \Gamma _{\mathbf{q},j} + \Gamma _{\mathbf{q},j'}}{(-\nu + \omega _{\mathbf{q},j} - \omega_{\mathbf{q},j'})^2+(\Gamma _{\mathbf{q},j} + \Gamma _{\mathbf{q},j'})^2}\right). \numberthis
    \label{pert_nondiag}
\end{align*}
This is a clean, easily interpreted formula. However, we had to make a very important approximation in Eq.~\ref{step_2}. Here we took limit $\nu \rightarrow (\omega _{\mathbf{q},j} - \omega _{\mathbf{q},j'})$ in terms outside of the Lorentzian. This breaks the limit $\nu \rightarrow 0$ and we can not recover the same expression for the DC coherent contribution we got in Ref.~\cite{GK-main}. For the particle contribution we do not have the same issue because in both cases we take the same limit $\nu\rightarrow 0$.

\newpage

\subsection{Static lattice thermal conductivity of CuCl}

\begin{figure}[h!]
    \centering
    \includegraphics[width=0.9\linewidth]{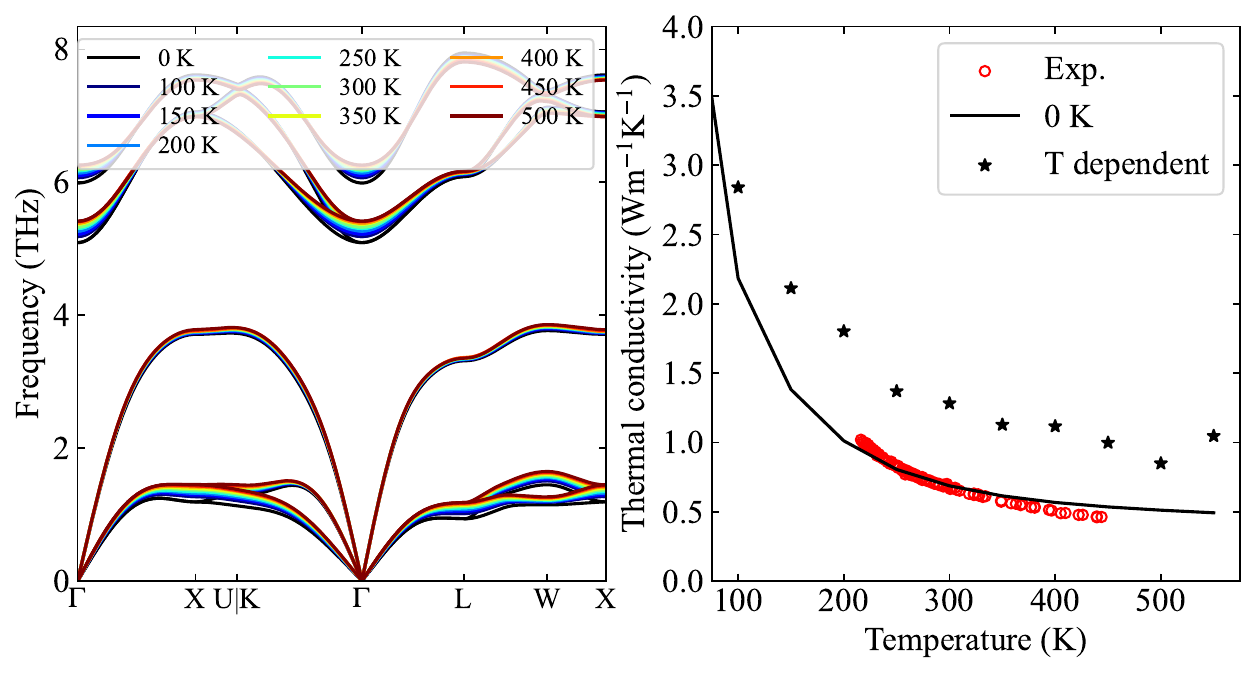}
    \caption{Left: Temperature dependent phonon dispersion of CuCl. Right: Lattice thermal conducitivity of CuCl calculated with 0 K force constants (black line) and temperature dependent force constants (black stars) compared with experiment (red circles).}
    \label{fig:comp_t}
\end{figure}

Previous first-principles calculations seriously overestimated lattice thermal conductivity of CuCl by almost factor of 2~\cite{CuClkappa_theory1, CuCl_kappa_theory2}. Our current calculations that include the SSCHA renormalized force constants at 0 K are in much better agreement with experiment at room temperature. This, however, is probably due to the fortuitous cancellation of errors. If we use the temperature dependent SSCHA force constants the agreement worsens, see Sup. Fig.~\ref{fig:comp_t}.

The phonon frequencies as calculated from SSCHA auxiliary force constants do not change significantly with temperature. However, the lattice thermal conductivity results do change significantly. The 0 K calculated force constants give much better agreement with experiment, while temperature dependent force constants overestimate it just like previous studies. From our previous studies we noticed that in most cases third order force constants soften with temperature which explains larger $\kappa$ with temperature dependent force constants~\cite{GK-main}.

The recent calculations including the fourth order scattering give very good agreement with experiment~\cite{cucl_4th_order}. Considering that they are calculating force constants by fitting Born-Oppenheimer energy surface the agreement points to strong influence of the 4th order anharmonicity. We assume that in our calculations, where we do not use 4th order scattering, this anharmonicity contribution is somewhat folded back to the 3rd order force constants calculated at 0 K which gives a better agreement with experiment. With increasing temperature 3rd order force constants soften and the inclusion of the 4th order anharmonicity is essential to quantitatively capture the lattice thermal conductivity in this material. Unfortunately, the scattering due to 4th order anharmonicity is still not fully implemented in SSCHA so we can not use it. 

It is worth to keep in mind that in our study we are more interested in studying the dynamical lattice thermal conductivity whose spectral dispersion depends more strongly on actual phonon frequency. These phonon frequencies do not change significantly with temperature as shown in Supp. Fig.~\ref{fig:comp_t}, meaning our results on dynamical lattice thermal conductivity are reliable and the discussion presented in the main part holds even without including 4th order anharmoncity. 

\newpage
\section{Dynamical lattice thermal conductivity of harmonic materials and comparison with mean free path approach}

\begin{figure}[h!]
    \centering
    \includegraphics[width=0.8\linewidth]{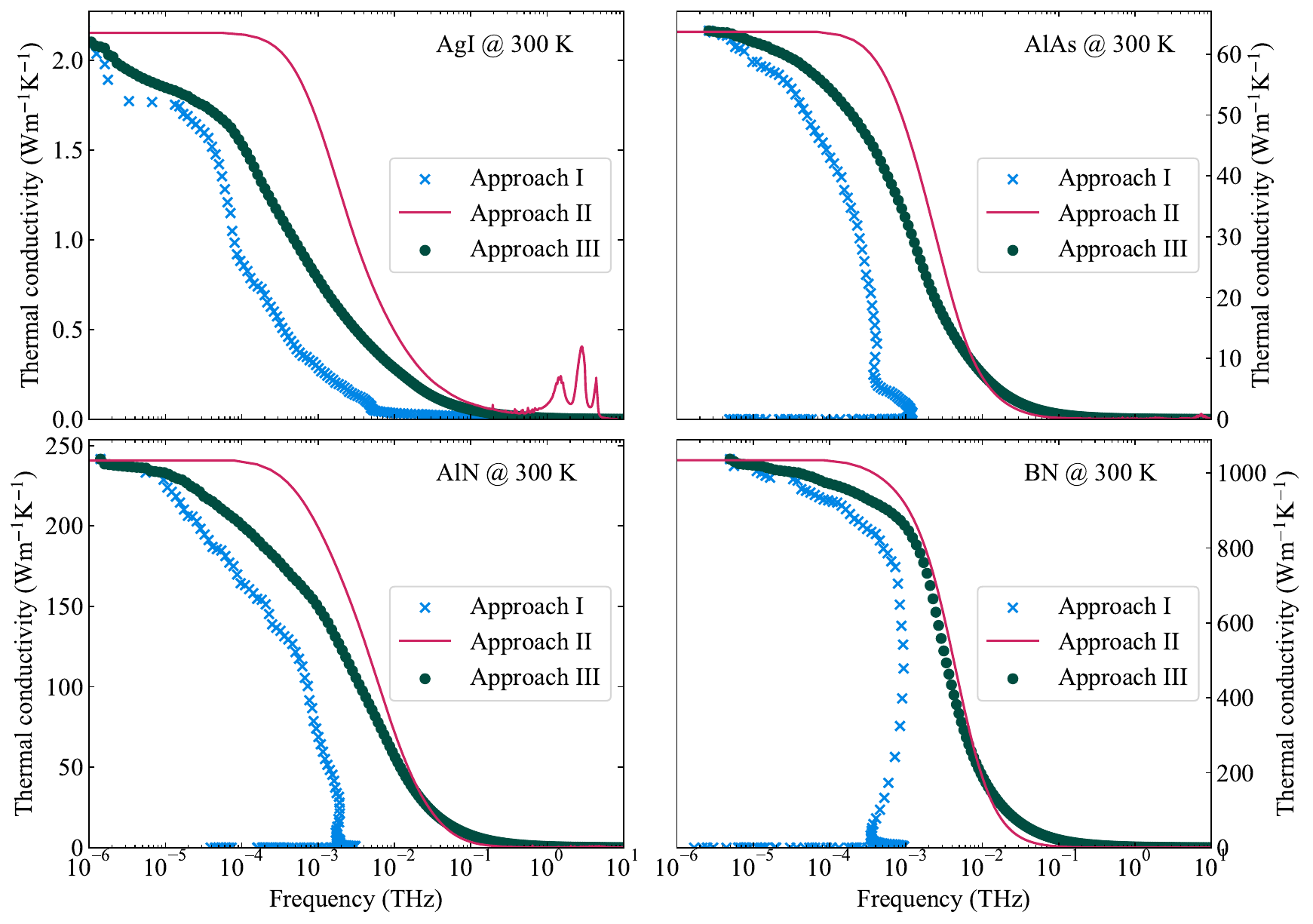}
    \caption{Dynamical lattice thermal conductivity of AgI, AlAs, AlN and BN at 300 K. Solid lines are calculations done using Supp. Eq.~\ref{diag_pert} and ~\ref{pert_nondiag}. Blue crosses are results for cumulative lattice thermal conductivity with mean free path converted to driving frequency $\nu = \frac{\kappa (L_p)}{\pi C L_p ^2}$. Green circles is total lattice thermal conductivity but with maximum phonon mean free path set to $L_p$. Force constants for these materials are taken form PhononDB database~\cite{CuCl_kappa_theory2}.}
    \label{fig:phonondb}
\end{figure}

The phenomenological explanation for the reduction of thermal conductivity with increasing frequency of the heat source is that heat has a finite penetration depth $L_p$: phonons with mean free paths ($l_{\mathbf{q},j}$) longer than this depth contribute little, leading to the observed drop in conductivity:
\begin{align*}
    \kappa ^{x,y} (L_p) = \frac{1}{NV}\sum _{\mathbf{q},j} \kappa ^{x,y} _{\mathbf{q},j}\Theta(L_p - l_{\mathbf{q},j}), \numberthis
    \label{cumm_kappa}
\end{align*}
Thus, the cumulative lattice thermal conductivity is taken as a proxy for the dynamical lattice thermal conductivity. However, from Eq.\ref{diag_pert} we see that the decay instead arises from the loss of coherence of phonon modes induced by the finite drive frequency. Casting Eq.\ref{diag_pert} in terms of phonon mean free paths $l_{\mathbf{q},j} = |\mathbf{v}_{\mathbf{q},j}|\tau _{\mathbf{q},j}$:
\begin{align*}
    \kappa (l_\nu) = \frac{1}{NV}\sum _{\mathbf{q},j}\frac{\kappa _{\mathbf{q},j}}{\left(\frac{l_{\mathbf{q},j}}{l_\nu}\right)^2 + 1}.
\end{align*}
Here $l_\nu = |\mathbf{v}_{\mathbf{q},j}|/\nu$ and should be the same as $L_p$ from the definition of the cummulative lattice thermal conductivity (Eq. 4 of the main part) if phenomenological model holds. If $l_{\mathbf{q},j} \gg l_\nu$, this phonon mode does not contribute significantly to the measured lattice thermal conductivity which agrees with the phenomenological explanation. However, for $l_{\mathbf{q},j} > l_\nu$ the contribution of the $(\mathbf{q},j)$ mode is reduced but not exactly 0. For this reason, we can not say that time dependent thermoreflectance experiments are measuring exactly cummulative lattice thermal conductivity and thus can not reliably be used to estimate phonon mean free paths. This is true even for very harmonic materials where most of the transport is particle-like and phonon mean free paths are relevant. In anharmonic materials, alloys, and amorphous solids where coherent contribution is significant the phenomenological explanation with phonon mean free paths can not be used at all. This we illustrate in Supp.Fig.~\ref{fig:phonondb}.

We propose a similar phenomenological explanation of the time dependent thermoreflectance experiments is to say that the maximum mean free path of the phonon mode is set to the heat penetration depth. In this case phonon modes with mean free paths larger than the penetration depth do not travel balistically, but rather scatter at the boundary defined by the penetration depth. 

These three approaches—(I) cumulative lattice thermal conductivity calculated assuming that phonons with a mean free path larger than a heat penetration depth do not contribute; (II) dynamical lattice thermal conductivity calculated with Supp. Eq.~\ref{diag_pert} and ~\ref{pert_nondiag}; and (III) total lattice thermal conductivity with the maximum mean free path set by the penetration depth—are shown in Supp. Fig.\ref{fig:phonondb}. In highly harmonic materials (BN in our example), approaches II and III agree almost perfectly, indicating that time-dependent thermoreflectance experiments can be interpreted as systems in which the maximum phonon mean free path is set by the heat penetration depth. As anharmonicity increases (or as $\kappa$ decreases), the agreement between approaches worsens, indicating that diffusive transport is not the only transport mode. The most anharmonic material in our dataset (AgI) shows the worst agreement between approaches. Also, only AgI shows noticeable resonant peaks in dynamical lattice thermal conductivity in THz range.

\end{document}